# Using SN Ia Light Curve Shapes to Measure The Hubble Constant

Adam G. Riess, William H. Press, Robert P. Kirshner

*Harvard-Smithsonian Center for Astrophysics, 60 Garden Street, Cambridge, MA 02138*

## Abstract

We present an empirical method which uses visual band light curve shapes (LCS) to estimate the luminosity of type Ia supernovae (SN Ia). This method is first applied to a "training set" of 8 SN Ia light curves with independent distance estimates to derive the correlation between the LCS and the luminosity. We employ a linear estimation algorithm of the type developed by Rybicki and Press (1992). The result is similar to that obtained by Hamuy et al. (1995a) with the advantage that LCS produces quantitative error estimates for the distance. We then examine the light curves for 13 SN Ia to determine the LCS distances of these supernovae. The Hubble diagram constructed using these LCS distances has a remarkably small dispersion of $\sigma_V = 0.21$ mag. We use the light curve of SN 1972E and the Cepheid distance to NGC 5253 to derive $67 \pm 7$ km s$^{-1}$ Mpc$^{-1}$ for the Hubble constant.

subject headings: supernovae:general ; cosmology: distance scale

# 1 SN Ia as Standard Candles

SN Ia are important as standard candles for cosmology because they are luminous ($10^9$ L$\odot$) and homogeneous ($\sigma_M$=0.65-0.36 in recent work). Yet, a handful of recent well-observed supernovae: SN 1986G (Phillips et al. 1987, Cristiani et al. 1992 ), SN 1991T (Phillips et al. 1992, Filippenko et al. 1992a), SN 1991bg (Leibundgut et al. 1993, Filippenko et al. 1992b ) and the comparison of SN 1992bc and SN 1992bo (Maza et al. 1994) provide convincing demonstrations that the SN Ia category includes objects of significantly different luminosities. Phillips (1993) (hereinafter P93) showed that the luminosity of SN Ia's is correlated with the decline in the 15 days following maximum light ($\Delta m_{15}$ ). Luminous SN Ia decline the least, and intrinsically dim SN Ia's decline most rapidly, as suggested earlier by Pskovskii (1984). This property of SN Ia light curves provides a way to estimate the intrinsic luminosity and to sharpen SN Ia's as tools for extragalactic astronomy. We analyze the variation in SN Ia light curve shapes and luminosities in a quantitative way that improves the utility of SN Ia's as distance indicators and leads to a value for H$_o$. We call this approach the LCS method (for Light Curve Shapes).

# 2 Basic Training

We apply linear estimation algorithms outlined by Rybicki and Press (1992) who developed tools to determine unknown linear parameters in a data set. We model a light curve as a sum of vectors. The coefficients of these vectors are a "luminosity correction" and the distance modulus, which can be determined by minimizing the residuals between the data and a model for the light curve. Here, we sketch the method: a subsequent paper will provide the formal details (Riess, Press, and Kirshner 1995).

We consider the vector **y** of measured apparent magnitudes to have the form

$$\mathbf{y} = \mathbf{s} + \mathbf{L}\mathbf{q} + \mathbf{n} \qquad (1)$$

Here **s**, the "signal", is the template light curve, in absolute magnitudes, of a SN Ia (Leibundgut 1988) and **n** is unmodeled and observational noise. The matrix **L** has columns which represent "correction" functions that can be added to the template to account for the variations among SN Ia light curves. The values in the vector **q** are the coefficients which describe how much of each correction function is added. For example, if **L** has a first column whose entries are unity, then the corresponding $q_1$ represents a fixed offset between the template (in absolute magnitude) and the observations (in apparent magnitude): the distance modulus. **L** can also have columns which are functions of time; these represent possible "correction templates" which, added in the right amount, correct the mean template to the shape of an observed light curve. For the simplest application to the supernova light curves **L** and **q** take the form:

$$\mathbf{L} = \begin{pmatrix} 1 & R(t_1) \\ 1 & R(t_2) \\ . & . \\ . & . \\ . & . \\ 1 & R(t_M) \end{pmatrix} \qquad \mathbf{q} = \begin{pmatrix} \mu_0 \\ \mu_1 \end{pmatrix} \qquad (2)$$

where $R(t)$ is a single correction template, $\mu_0$ the distance modulus, and $\mu_1$ the luminosity correction defined relative to template-like SN Ia's. We assume that the amount by which the supernova light curve shape deviates from the template is proportional to the difference in absolute magnitude.

The formula for the $\chi^2$ of the fit between model and data is:

$$\chi^2 = (\mathbf{y} - \mathbf{s} - \mathbf{Lq})^{\mathbf{T}} \ \mathbf{C}^{-1} \ (\mathbf{y} - \mathbf{s} - \mathbf{Lq}) \qquad (3)$$

where **C** is the noise correlation matrix whose elements are intended to reflect the errors in observations and residual, unmodeled correlations in SN Ia light curves. The inverse of **C** contains the weights of observations in a model reconstruction. We perform two analytic minimizations of $\chi^2$ in equation (3). One is the best estimate of **L** provided **q** is known, and the other is the best estimate of **q** when **L** is known.

We use a "training set" of SN Ia light curves to find the best estimate for the correction template contained in **L**, because for this set we know the entries of **q** which are the distance moduli and the luminosity corrections. For training we use the P93 set of 1980N, 1981B, 1986G, 1989B, 1990N, 1991T, 1991bg, and 1992A all of which have precise optical photometry, well-sampled light curves, and accurate relative distances to the host galaxies based on surface brightness fluctuations or the luminosity-line width relation. It is convenient to choose the "normal" peak luminosity as the average of SN 1980N, 1981B, and 1989B, and our "luminosity corrections" give the difference in absolute magnitude between a particular SN Ia and this typical set.

For **L** in the form of equation (2), minimization of $\chi^2$ in equation (3) with respect to $R(t)$, which turns out to be mathematically independent of **C**, yields;

$$R(t - t_p) = \frac{\langle [m(t) - M(t - t_p) - \mu_0]\mu_1 \rangle}{\langle \mu_1^2 \rangle} \qquad (4)$$

where $m(t)$ represents apparent magnitudes for a training set light curve, $\mu_0$ is the distance modulus, $M(t - t_p)$ is the usual template (Leibundgut 1988) as a function of time past the peak, and $\mu_1$ is the luminosity correction defined as the absolute magnitude difference between the supernova maximum and the normal (template) peak. The angle brackets denote an average over the training set, where each supernova is weighted by the stated uncertainty in the distance estimate to its galaxy. Leibundgut's template approximates the light curves of the "normal" supernovae in the training set very well, though the method is not sensitive to details of the template. Various amounts of the function R(t) added to the template, provide useful reconstructions of the full range of light curves in the training set, as illustrated in Figure 1.

With **L** determined from the training set, we minimize $\chi^2$ in equation (3) with respect to **q** to best estimate the distance modulus and the luminosity correction of other SN Ia *solely from the observed light curve*. This procedure estimates the distance modulus and the luminosity correction (the latter now a "nuisance" parameter) simultaneously *over the entire light curve* reducing the dependence on any single, critical time and decreasing our propagated error. However, most of the information that is useful for estimating the distance modulus comes from observations within three weeks of maximum light.

We estimate the uncertainty of the parameters in **q** from equation (3) by searching the error ellipsoid for the one sigma confidence interval. These errors reflect how the $\chi^2$ of the fit between the data and the light curve reconstruction constrain the estimates of peak time, luminosity correction, and distance modulus. These uncertainties are strongly tied to the light curve sampling, measuring errors, and conformity to our linear model as well as the accuracy of the correction template garnered from the training set.

## 3  The Hubble Diagram for an Independent Sample of SN Ia

Mario Hamuy and his collaborators in the Calán/Tololo Supernova Search have generously allowed us to analyze their data in advance of publication. (Hamuy et al. 1993a,1994,1995a,1995b) Utilizing these 10 SN Ia's (SN 1992bo, SN 1992bc, SN 1992K, SN 1992aq, SN 1992ae, SN 1992P, SN 1992J, SN 1991U, SN 1991ag, SN 1990af) plus SN 1993ae from our own work (Riess et al. 1995) and 2 from the literature (SN 1992G and SN 1991M, observed by Ford et al. 1993), we find the best **q** for each supernova using the **L** matrix derived from the training set. All of these supernovae have observations which begin ∼10 days after V maximum, which provides a reliable LCS estimate.

For each supernova, we minimize the $\chi^2$ of equation (3) with respect to three parameters: two components of **q** (which can be found by linear algebra) and the time of the maximum (which enters nonlinearly, so must be found by an outer iteration). For every choice of the time of maximum, proper K corrections and $(1+z)$ time dilation were applied (Hamuy et al. 1993b) to return the light curve to the rest frame.

If LCS has improved the distance modulus estimate for each of these new supernovae, it should decrease the scatter in a Hubble diagram when compared to a Hubble diagram that uses the assumption that all SN Ia have the same luminosity. To test this, we construct a Hubble diagram assuming a standard luminosity and a standard template (Leibundgut 1988) for fitting the maximum, without using information from the light curve shape. The result is shown in Figure 2a.

The dispersion in the Hubble diagram is $\sigma_m = 0.50$, a typical value when no "peculiar" SN Ia's have been discarded (Barbon, Capaccioli, and Ciatti 1975, Branch and Bettis 1978, Branch 1982, Tammann 1982, Sandage and Tammann 1985, Tammann and Leibundgut 1990, Della Valle and Panagia 1992, Sandage and Tammann 1993).

Next, we conduct a direct comparison with the LCS method. The result is Figure 2b, which shows a remarkably small scatter: the dispersion is $\sigma_m = 0.21$, a reduction in the scatter by a factor of 2.4, without discarding any objects because of spectroscopic or photometric properties. This dramatic reduction in the dispersion transforms SN Ia into one of the best available distance indicators.

Although our analysis started with the distance scale of P93, (where surface brightness fluctuations and line widths are the distance indicators), we only employ the relative distances in our training. To determine the Hubble constant, we use the HST Cepheid distance to NGC 5253 (Sandage et al. 1994, Saha et al. 1994) which contained the well-observed SN 1972E (Lee et al. 1972, Ardeberg et al. 1973, Kirshner et al. 1973). Fitting the standard template to the light curve and assuming this supernova has the same absolute brightness as the typical object in the sample yields $H_o = 53 \pm 11$ km s$^{-1}$ Mpc$^{-1}$ from figure 2a. This is consistent with the result of Sandage and Tammann (1993), Sandage et al. (1994) who estimated $H_o = 58 \pm 9$ km s$^{-1}$ Mpc$^{-1}$ (internal error only) using SN 1972E as a standard candle and a completely independent set of supernova light curves.

When we use the LCS method on SN 1972E, as for the other 13 SN Ia in our sample, we find SN 1972E was 0.24 mag brighter than the template, which yields $H_o = 67$ km s$^{-1}$ Mpc$^{-1}$ from the intercept of Figure 2b. Half of the increase in the Hubble constant comes from the excess luminosity of SN 1972E, and the other half from a shift in the ridge line produced by applying LCS to the sample.

What is the effect of reddening on the predicted parameters? We apply corrections for absorption in our galaxy (Burstein & Heiles 1982), but none for the internal absorption of the host galaxies. It is likely that some absorption affects the supernovae in the galaxies of the new sample. However, Sandage and Tammann (1993) find that the dispersion in their Hubble diagram *increases* when corrections for extinction are made by assuming a SN Ia color at maximum. Current work suggests that the color at maximum may be correlated with other supernova parameters including luminosity (P93, Wells et al. 1994). Rather than introduce an uncertain estimate for the extinction, we leave out corrections for host galaxy absorption in the sample and for 1972E, which does not bias our Hubble constant.

What is the error attached to the estimate of $H_o$? The uncertainty in an uncorrected SN Ia is found to be 0.39 mag by requiring $\chi^2 = 12$ for the 12 degrees of freedom (13 supernovae minus one fitted degree for $H_o$) in the Hubble diagram. Applied to SN 1972E this error provides the largest source of uncertainty for $H_o$. By comparison, the LCS method predicts individual error estimates for each light curve (as described in §2), and the validity of these error estimates is substantiated by the resulting value of $\chi^2 = 10$ for Figure 2b. When we apply the LCS method to SN 1972E, the error estimate for the distance modulus of NGC 5253 is 0.08 mag, so the largest uncertainty becomes the zero point of the Cepheid scale, uncertain by 0.15 mag (Feast and Walker 1987). Compounding the one sigma confidence interval in the fit to the ridge line, the uncertainty in the Cepheid measurements for SN 1972E, and 0.10 mag for the possible difference in extinction between SN 1972E and the mean of our sample we get an overall uncertainty in the LCS estimate of $H_o$. Our best estimate for the Hubble constant using LCS to measure the distances is $H_o = 67 \pm 7$ km s$^{-1}$ Mpc$^{-1}$, compared to $H_o = 53 \pm 11$ km s$^{-1}$ Mpc$^{-1}$ for the standard candle version.

Our analysis is independent of the recent and similar template-fitting approach developed by Hamuy et al. (1995a). Since we employ a sample of SN Ia's that partially overlaps with theirs (7 of 13 objects) and we use the same light curve (SN 1972E) to establish the zero-point, it is reassuring to find our estimate of $H_o$ is in close agreement with their estimate of $H_o = 62-67$ km s$^{-1}$ Mpc$^{-1}$. Another approach to measuring the $H_o$ from supernovae, the Expanding Photosphere Method for SN II, gives a consistent value of $73 \pm 7$ km s$^{-1}$ Mpc$^{-1}$ (Schmidt, Kirshner and Eastman 1994). Since that method is completely independent of the Cepheid distance scale, and the SN Ia method depends on it, the agreement augurs well for further resolution of the distance scale.

## 4 Discussion

Our distance estimates are robust, but there are limits to the quality of the light curves from which we can make accurate predictions. We do not use photographic light curves to avoid introducing significant systematic errors from poor background subtraction, mixing modern and ancient filter functions, and mistaking inconsistent photometry for real intrinsic deviation in the light curve. We omit light curves which begin more than ten days after maximum (as determined by our template fit) because their limited photometric history does not predict the luminosity.

Despite its modest underpinnings, LCS produces successful estimates for supernova distances. When we applied the results of the training set to a new group of SN Ia's with $\sigma_{m_V} = 0.50$ mag, using LCS reduced their dispersion to $\sigma_m = 0.21$ mag. The remaining dispersion is likely a result of peculiar velocities, absorption, uncertain K corrections, photometry errors, and unmodeled behavior. Overall, LCS makes a custom reproduction of each light curve while solving for distance, luminosity correction from the template, and the time of maximum, with errors on each of these parameters.

Our analysis is completely empirical, yet it is inviting to speculate on the supernova property which produces a one parameter family of light curves: the obvious candidate is mass, either as total mass or nickel mass yield. Comparing our family of empirical light curves with computed light curves for various models, may determine which mechanisms for SN Ia's best match the facts.

We are grateful to Mario Hamuy, Mark Phillips, Nick Suntzeff, and the entire Calán/Tololo collaboration for the opportunity to study their outstanding data before publication. We have been assisted by George Rybicki (who knows how to complete the square) and Brian Schmidt (who knows his way to the Square). This work was suppported through grants AST-92-18475 and PHY-91-06678.

# References

Ardeberg, A.L., de Groot, M.J. 1973, A&A, 28, 295

Barbon, R., Capaccioli, M., & Ciatti, F. 1975, A&A, 44, 267

Branch, D., & Bettis, C. 1978, AJ, 83, 224

Branch, D. 1982, ApJ, 258, 35

Burstein, D., & Heiles, C. 1982, AJ, 87, 1165

Cristiani, S. et al 1992, A&A, 259,63

Della Valle, M., & Panagia, N. 1992, AJ, 104,696

Feast, M.W., & Walker, A.R. 1987, ARA&A, 25,345

Filippenko, A.V. et al 1992a, ApJ, 384, L15

Filippenko, A.V. et al 1992b, AJ, 104, 1543

Ford, C. et al 1993, AJ, 106, 3

Hamuy, M., Phillips, M.M., Maza, J., Suntzeff, N.B., Schommer, R.A., & Aviles, A. 1995b, in preparation

Hamuy, M., Phillips, M.M., Maza, J., Suntzeff, N.B., Schommer, R.A., & Aviles, A. 1995a, AJ, in press

Hamuy, M., Phillips, M.M., Maza, J., Suntzeff, N.B., Schommer, R.A., & Aviles, A. 1994, AJ, in press

Hamuy, M., et al 1993a, AJ, 106, 2392

Hamuy, M., Phillips, M., Wells, L., & Maza, J. 1993b, PASP, 105, 787

Kirshner, R.P., Oke, J.B., Penston, M.V., & Searle, L. 1973, ApJ, 185, 303

Lee, T.A., Wamsteker, W., Wisniewski, W.Z., & Wdowiak, T.J. 1972, AJ, 177, L59

Leibundgut, B. 1988, PhD Thesis, Basel

Leibundgut, B. et al 1993, AJ, 105, 301

Maza, J., Hamuy, M., Phillips, M., Suntzeff, N., Aviles, R. 1994, ApJ, 424, L107

Phillips, M. et al 1987, PASP 99,592

Phillips, M. 1993, ApJ, 413, L105

Phillips, M., Wells, L., Suntzeff, N., Hamuy, M., Leibundgut, B., Kirshner, R. P., Foltz, C., 1992, AJ, 103, 1632

Pskovskii, Y. 1984, Sov. Astron., 28,658

Riess, A.G., Press W.H., Kirshner, R.P., 1995 (in preparation)

Riess, et al, 1995 (in preparation)

Rybicki, G. & Press, W., 1992,ApJ, 398,169

Saha, A., Labhardt, L., Schwengeler, H., Macchetto, F.D., Panagia, N., Sandage, A., & Tammann, G. A. 1994, ApJ, in press

Sandage, A. et al 1994, ApJ, 423, L13

Sandage, A., & Tammann, G.A. 1985, in *Supernovae as Distance Indicators*, eds. Bartel, N., Springer:Berlin

Sandage, A., & Tammann, G.A. 1993, ApJ 415,1

Schmidt, B.P., Kirshner, R.P., & Eastman, R. 1994, ApJ, 000, 000

Tammann, G.A. 1982, in *Supernovae: A Survey of Current Research*, eds. Rees, M.J., Stoneham, R. J., Dordrecht:Reidel

Tammann, G.A., & Leibundgut, B. 1990, A&A

Wells, L. et al 1994, AJ (in press)

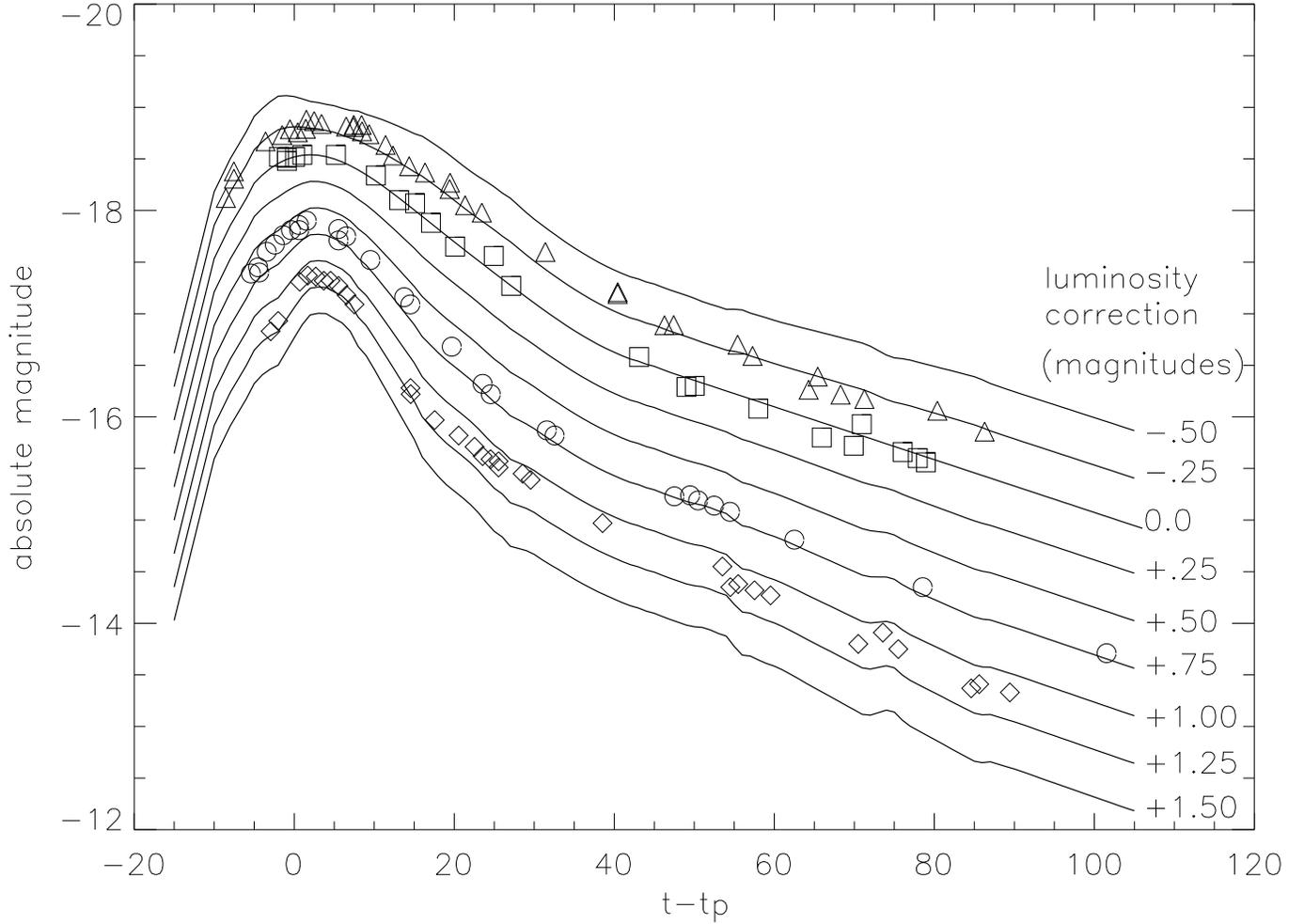

Figure 1: Empirical family of visual band SN Ia light curves. This sample of empirical light curves is derived from the training set and depicts the entire range of light curve shapes and their correlation with luminosity (on the P93 distance scale). This set is obtained by adding the correction template, R(t-$t_p$), multiplied by various luminosity corrections, $\mu_1$, to the normal template and allows for the best reconstruction of a light curve and its distance modulus. Data shown as reconstructed, 91T=△, 81B=□, 86G=○, 91bg=◇

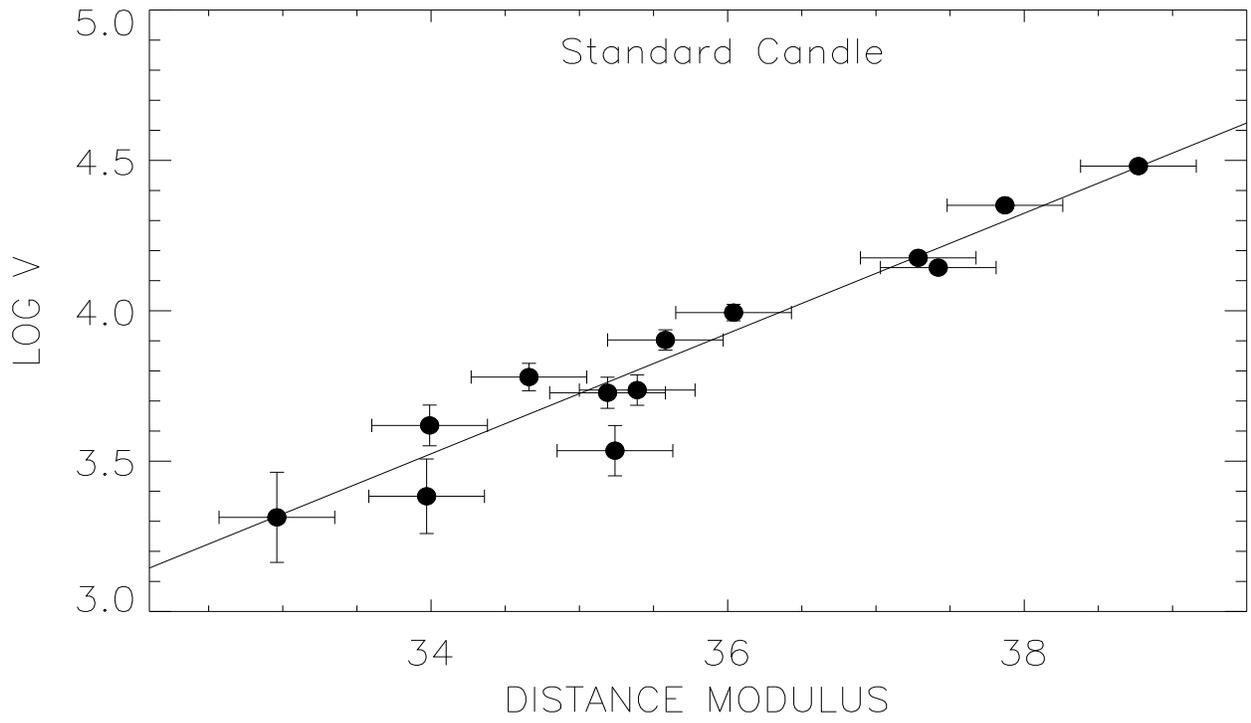

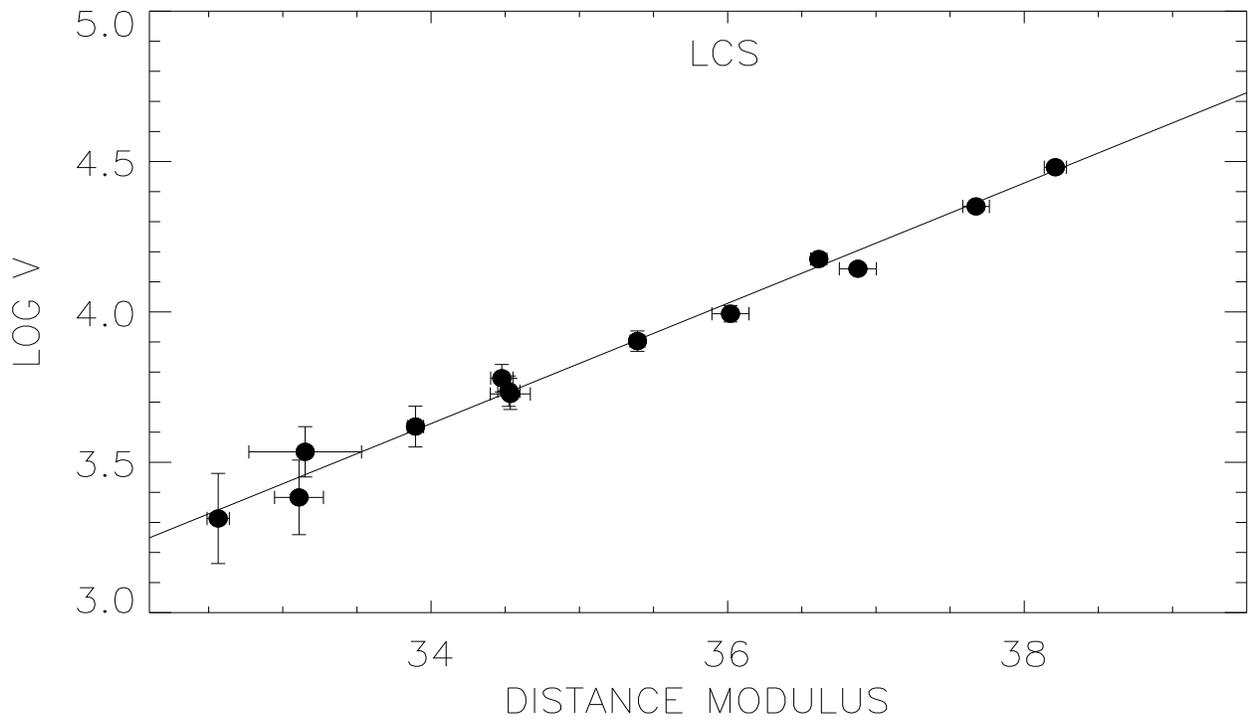

Figure 2: Hubble Diagrams for SN Ia with velocities in the COBE rest frame and distances calibrated with 1972E. All velocity errors are 600 km s$^{-1}$ reflecting a plausible estimate of random velocities with respect to the Hubble flow. (a) Distances estimated with a standard luminosity and a standard template light curve. This method yields $\sigma_v$=0.50 and $H_0$=53 ± 11 km s$^{-1}$ Mpc$^{-1}$ (b) Distances from the LCS method. This method yields $\sigma_v$=0.21 and $H_0$=67 ± 7 km s$^{-1}$ Mpc$^{-1}$.